\documentclass[ 12pt, reprint,
showpacs,
nofootinbib,
floatfix,
 amsmath,amssymb,
 aps,
]{revtex4-1} 
\usepackage{epsfig}
\usepackage{amsmath}
\usepackage{amsfonts}
\usepackage{amssymb}
\usepackage{color}
\usepackage{bm}
\usepackage{dcolumn}
\usepackage{rotating}

\newcommand{\be}{\begin{equation}}
\newcommand{\ee}{\end{equation}}
\newcommand{\bea}{\begin{eqnarray}}
\newcommand{\eea}{\end{eqnarray}}

\newcommand{\mbss}[1]{_{\mbox{\scriptsize #1}}}

\newcommand{\mbsu}[1]{\mbox{\scriptsize #1}}

\newcommand{\vphu}{\vphantom{*}}
\newcommand{\vphd}{\vphantom{1}}

\newcommand{\ve}{\varepsilon}

\definecolor{dgreen}{rgb}{0.0,0.5,0.0}

\begin{document}

\title{Optimizing phonon space in the phonon-coupling model}

\author{V. Tselyaev}
\author{N. Lyutorovich}
\affiliation{St. Petersburg State University, St. Petersburg, 199034, Russia}
\author{J. Speth}
\email{J.Speth@fz-juelich.de}
\affiliation{Institut f\"ur Kernphysik, Forschungszentrum J\"ulich,
D-52425 J\"ulich, Germany}
\author{P.-G. Reinhard}
\affiliation{Institut f\"ur Theoretische Physik II,
Universit\"at Erlangen-N\"urnberg,
D-91058 Erlangen, Germany}

\date{\today}

\begin{abstract}
We present a new scheme to select the most relevant phonons in the
phonon-coupling model, named here time-blocking approximation (TBA).
The new criterion, based on the phonon-nucleon coupling strengths
rather than on $B(EL)$ values, is more selective and thus produces
much smaller phonon spaces in TBA. This is beneficial in two respects:
first, it curbs down the computational cost, and second, it reduces
the danger of double counting in the expansion basis of TBA.  We use
here TBA in a form where the coupling strength is regularized to keep
the given Hartree-Fock ground state stable.  The scheme is implemented
in an RPA and TBA code based on the Skyrme energy functional. We first
explore carefully the cutoff dependence with the new criterion and can
work out a natural (optimal) cutoff parameter. Then we use the freshly
developed and tested scheme to a survey of giant resonances and
low-lying collective states in six doubly magic nuclei looking also on
the dependence of the results when varying the Skyrme parametrization.
\end{abstract}

\pacs{21.10.Re, 21.60.Jz, 24.10.Cn, 24.30.Cz}

\maketitle

\section{Introduction}
\label{introduction}

The most often applied approach to collective phenomena in nuclear
physics is the \emph{random phase approximation} (RPA) \cite{Boh53a}.
It provides a reliable description of the spectral distribution
of multipole excitations from low energy collective states to the
giant resonance region. If one includes the single-particle ($sp$)
continuum we end up with the \emph{continuum RPA} {\cite{Krewald_1974}}
which allows to calculate the escape width of giant resonances.
In medium and heavy mass nuclei the experimental width however is
dominated by the spreading width {which involves} higher order
effects. These may be explicitly {two-particle two-hole} 
correlations \cite{Drozdz_1990} or the coupling to phonons which also
give rise to a broadening 
\cite{Soloviev_1976,Krewald_1977,Tselyaev_1989,Kamerdzhiev_1993,Col01a}
of the strength. The most complete formulation of the
quasiparticle-phonon coupling has been developed in
Ref. \cite{Tselyaev_2007} which is the basis of the present
investigation. This approach is called \emph{quasiparticle time
blocking approximation} (TBA). As input one needs $sp$ energies, 
$sp$ wavefunctions and a particle-hole ($ph$)
force. Here we have to distinguished two different approaches.
In the first case the $sp$ quantities are taken from a shell model which
parameters are adjusted to the experimental $sp$ energies.
The $ph$ force is, e.g., modeled as zero-range, density dependent
interaction and the corresponding parameters are adjusted
to appropriate nuclear structure properties
(Refs. \cite{Kamerdzhiev_2004,Litvinova_2007}). 
In the second case, one starts from an effective Lagrangian or Hamiltonian
which allows a fully self-consistent description of nuclear ground
state and subsequent dynamics, see Ref. \cite{Tselyaev_2016} and
references therein. Both approaches can equally well be complemented
by TBA to account for complex configuration.

As compared to other treatments of complex configurations
(e.g. second RPA \cite{Spe91b}), the phonon-coupling method
(i.e. TBA) is particularly efficient by confining the complex
configurations to a well manageable amount of phonons.
Thus the key task of TBA is to select properly the most relevant
phonons (i.e. RPA modes). It is obvious that we should chose those
phonons which incorporate a large amount of the interaction.
These are the collective phonons which consist of a coherent superposition
of many {$ph$} states. Besides delivering the strongest contributions,
using collective phonons is not so much plagued by double counting
as predominantly sp excitations do.
In the present paper we present a new method of selecting the phonon
space for TBA which is applicable to light and heavy
nuclei. This method is presented in {section \ref{newcriterion}} and
compared with {previously} used selection criteria.

\begin{figure}
\centerline{
\begin{picture}(80,65)
\put(0,0){
\includegraphics[width=0.9\linewidth,angle=0]{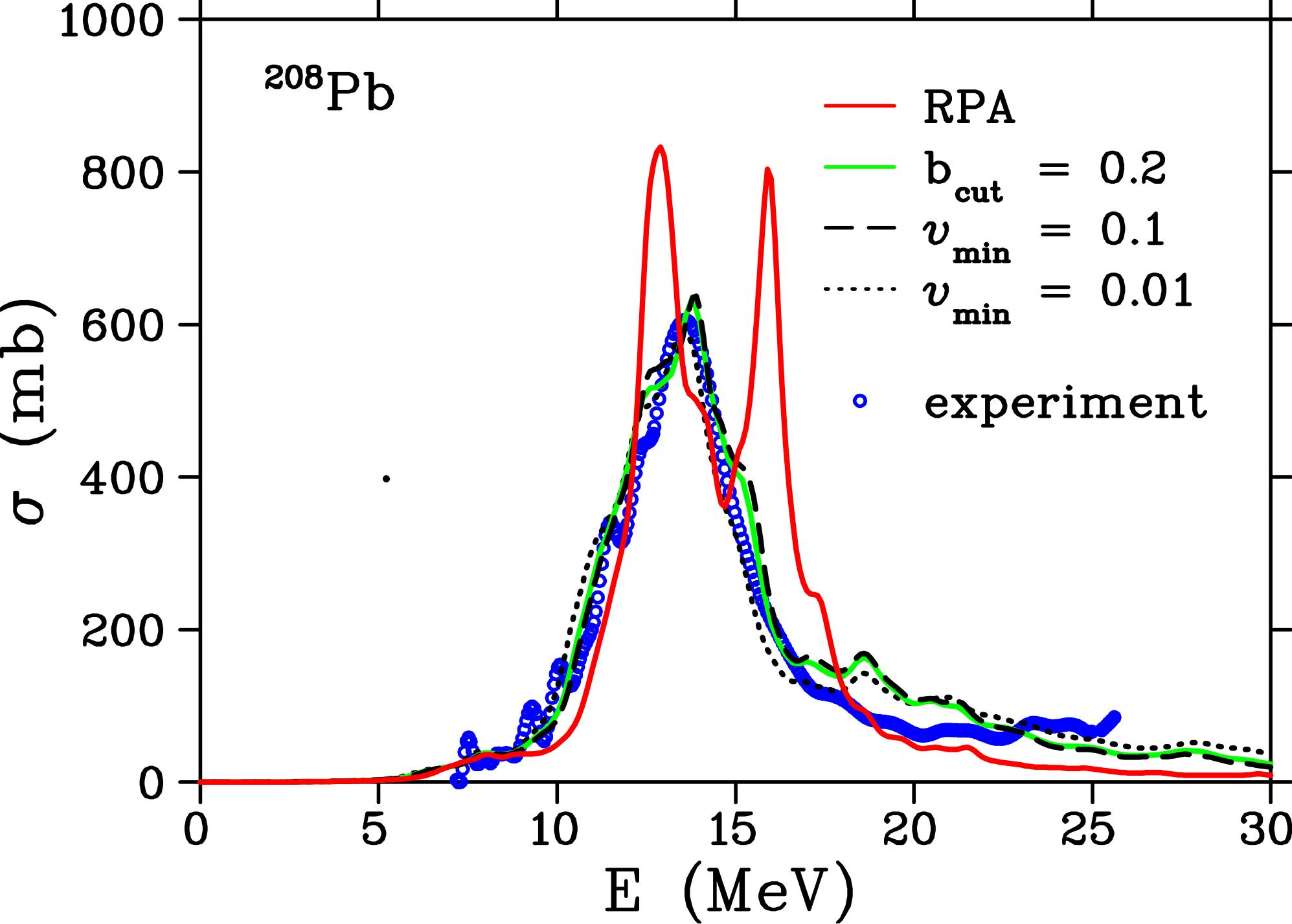}}
\end{picture}
}
\caption{\label{pb208} (Color online) The giant dipole resonance in
$^{208}$Pb calculated using RPA (red solid line) and the TBA with
different selection criteria for phonons: $B$-criterion with
$b_{\mbsu{cut}}=0.2$ (green solid line) and $V/E$-criterion with
$v_{\mbsu{min}}=0.1$ (black dashed line) and $v_{\mbsu{min}}=0.01$
(black dotted line). The Skyrme parametrization SV-bas was
used. Details of the methods and numerical parameters are described
in Sec.~\ref{newcriterion}.
Experimental data from \cite{Belyaev_1995}. }
\end{figure}
As a preview, figure \ref{pb208} illustrates the effect of TBA and the
impact of phonon space for the example of isovector dipole strength in
$^{208}$Pb (details of the method will be explained later). The RPA
result resides correctly in the region of the giant dipole resonance
(GDR), but has a marked double-peak structure which is at variance
with data. The spreading width described with TBA dissolves the upper
peak through substantial broadening and so recovers nicely the
experimental one-peak structure. TBA results are shown for three
different choices of phonon space.  Formerly, we used the $B(EL)$
strength as measure of collectivity taking into account only phonons
above a certain cutoff value. The curve marked $b_\mathrm{cut}=0.2$
stems from this old recipe. In this paper, we will present a new and
more selective criterion relying on the {phonon-nucleon coupling.
We denote this method the $V/E$-criterion and the corresponding
cutoff parameter $v_{\mbss{min}}$.} The curve $v_{\mbss{min}}=0.1$
yields practically the same results as previously, however, employing
much less phonons. The dependence on cutoff is indicated with the
curve $v_{\mbss{min}}=0.01$ which makes a slight difference as
compared to $v_{\mbss{min}}=0.1$.  In the following, we will discuss
in detail the optimal choice of the cutoff parameter.

Section \ref{newcriterion} starts in subsections \ref{sec:SumRPA}
  with a brief summary of RPA, in subsection \ref{sec:SumTBA} of TBA,
  and explains in subsection \ref{sec:select} the criteria for
  selecting the space of most relevant phonons. The latter subsection
  is decisive as it provides the formal basis for the new selection
  criterion from which we show later on that it is more efficient than
  previous choices.
Section \ref{detailscalculation} present details of the practical
treatment, the continuum-response formalism in subsection
\ref{sec:basicEq}, numerical aspects in subsection \ref{sec:numer},
and the actual nuclear mean field related to the Skyrme energy-density
functional in subsection \ref{sec:skyrme}.
Section \ref{dependencies} addresses the central question, namely the
dependence of the TBA results on the selection of phonons.  In
subsection A the previously used $B$-criterion which relies on the
magnitude of the $B(EL)$ values is compared with the new
$V/E$-criterion and in subsection B we discuss the dependence on the
new cutoff parameter $v_{\mbss{min}}$.  Here we demonstrate in
several figures that this new selection criterion gives rise to a
plateau in nearly all cases.  Surprisingly also the energies of the
low-lying collective $3^-$ resonances, which are the most sensitive
quantities in this respect, depend only smoothly on the new
parameter. This is connected with the stability conditions
\cite{Tselyaev_2013} introduced into the TBA approach. In section
\ref{GR}, we present our results for the giant multipole resonances,
giant monopole resonance (GMR), GDR and giant quadrupole resonance (GQR)
for light, medium and heavy mass nuclei.
Three Skyrme parameter sets were used with different effective masses.
In section \ref{lowlying}, the excitation energies and transition probabilities
for the low-lying collective $3^-$ and $2^+$ states are compared with
the experimental values. In both cases, the influence of the phonons
is discussed. Finally in section \ref{summary}, we summarize the paper
and give an outlook on further improvements.

\section{New criterion for the selection of the phonons}
\label{newcriterion}

\subsection{Summary of RPA}
\label{sec:SumRPA}

The RPA modes $n$ are characterized by the energies
$\omega^{\vphu}_{n}$ and transition amplitudes $z^{n}_{12}$ which
describe the composition of $n$ from the {$ph$ and $hp$} states $12$.
These are determined by the eigenvalue equation
\be
\sum_{34} \Omega^{\mbss{RPA}}_{12,34}\,z^{n}_{34} =
\omega^{\vphu}_n\,z^{n}_{12}\,,
\label{rpaze}
\ee
where
\be
\Omega^{\mbss{RPA}}_{12,34} = \Omega^{(0)}_{12,34}
+ \sum_{56} M^{\mbss{RPA}}_{12,56}\,{V}^{\vphu}_{56,34}\,,
\label{omrpa}
\ee
\be
\Omega^{(0)}_{12,34} =
h^{\vphu}_{13}\,\delta^{\vphu}_{42} -
\delta^{\vphu}_{13}\,h^{\vphu}_{42}\,,
\label{omrpaz}
\ee
\be
M^{\mbss{RPA}}_{12,34} =
\delta^{\vphu}_{13}\,\rho^{\vphu}_{42} -
\rho^{\vphu}_{13}\,\delta^{\vphu}_{42}\,,
\label{mrpa}
\ee
$\rho$ is the single-particle density matrix,
$h$ is the single-particle Hamiltonian, and
$V$ is the amplitude of the residual interaction.
In symbolic notation, Eqs. (\ref{rpaze}) and (\ref{omrpa}) read
\be
\Omega^{\mbss{RPA}}_{\vphd} |\,{z}^{n} \rangle =
\omega^{\vphu}_{n}\,|\,{z}^{n} \rangle\,,
\label{rpazes}
\ee
\be
\Omega^{\mbss{RPA}}_{\vphd} = \Omega^{(0)}_{\vphd}
+ M^{\mbss{RPA}}_{\vphd}V.
\label{omrpas}
\ee
The transition amplitudes $|\,{z}^{n} \rangle$ are normalized to
\be
\langle\,{z}^{n}\,|\,M^{\mbss{RPA}}_{\vphd}
|\,{z}^{n'} \rangle =
\mbox{sgn}(\omega^{\vphu}_{n})\,\delta^{\vphu}_{n,\,n'}.
\label{zmz}
\ee
We will suppose that the RPA is self-consistent (though this is not
essential for the subsequent formulas), i.e. that the following relations
are fulfilled:
\be
h^{\vphu}_{12} = \frac{\delta E[\rho]}{\delta\rho^{\vphu}_{21}}\,,\qquad
{V}^{\vphu}_{12,34} =
\frac{\delta^2 E[\rho]}
{\delta\rho^{\vphu}_{21}\,\delta\rho^{\vphu}_{34}}\,,
\label{sccond}
\ee
where $E[\rho]$ is an energy density functional.
In the basis diagonalizing the operators $h$ and $\rho$
we have:
\be
h^{\vphu}_{12} = \ve^{\vphu}_{1}\delta^{\vphu}_{12}\,,
\qquad
\rho^{\vphu}_{12} = n^{\vphu}_{1}\delta^{\vphu}_{12}\,,
\label{spbas}
\ee
where $n^{\vphu}_{1}=0,1$ is the occupation number.  In what follows
the indices $p$ and $h$ will be used to label the single-particle
states of the particles ($n^{\vphu}_{p} = 0$) and holes
($n^{\vphu}_{h} = 1$) in this basis.

\subsection{Summary of TBA}
\label{sec:SumTBA}

In TBA, Eq.~(\ref{rpaze}) takes the form
\be
\sum_{34} \Omega^{\mbss{TBA}}_{12,34}
(\omega^{\vphu}_{\nu})\,z^{\nu}_{34} =
\omega^{\vphu}_{\nu}\,z^{\nu}_{12}\,,
\label{tbaze}
\ee
where
\be
\Omega^{\mbss{TBA}}_{12,34}(\omega) = \Omega^{\mbss{RPA}}_{12,34}
+ \sum_{56} M^{\mbss{RPA}}_{12,56}\,\bar{W}^{\vphu}_{56,34}(\omega)\,,
\label{omtba}
\ee
\be
\bar{W}^{\vphu}_{12,34}(\omega) =
{W}^{\vphu}_{12,34}(\omega) - {W}^{\vphu}_{12,34}(0)\,.
\label{wsub}
\ee
The matrix ${W}(\omega)$ in (\ref{omtba}) and (\ref{wsub}) represents
the induced interaction and is defined in the $ph$ subspace as
\be
{W}^{\vphu}_{12,34}(\omega) =
\sum_{c,\;\sigma}\,\frac{\sigma\,
{F}^{c(\sigma)}_{12}
{F}^{c(\sigma)*}_{34}}
{\omega - \sigma\,\Omega^{\vphu}_{c}}\,,
\label{wdef}
\ee
where $\sigma = \pm 1$, $\,c = \{p',h',n\}$ is an index of the subspace
of $ph\otimes$phonon configurations, $n$ is the phonon's index,
\be
\Omega^{\vphu}_{c} =
\ve^{\vphu}_{p'} - \ve^{\vphu}_{h'} + \omega^{\vphu}_{n}\,,
\quad \omega^{\vphu}_{n}>0\,,
\label{omcdef}
\ee
\be
{F}^{c(-)}_{12}={F}^{c(+)*}_{21},\qquad
{F}^{c(-)}_{ph}={F}^{c(+)}_{hp}=0\,,
\label{fcrel}
\ee
\be
{F}^{c(+)}_{ph} =
\delta^{\vphu}_{pp'}\,g^{n}_{h'h} -
\delta^{\vphu}_{h'h}\,g^{n}_{pp'},
\label{fcdef}
\ee
$g^{n}_{12}$ is an amplitude of the quasiparticle-phonon interaction.
These $g$ amplitudes (along with the phonon's energies
$\omega^{\vphu}_{n}$) are determined by the positive frequency
solutions of the RPA equations as
\be
g^{n}_{12} = \sum_{34} {V}^{\vphu}_{12,34}\,z^{n}_{34}\,.
\label{gndef}
\ee

It is important to note the subtraction of the zero-frequency
interaction in the induced interaction $\bar{W}(\omega)$.  This serves
to confine the induced interaction only to dynamical excitations while
the ground state remains unaffected \cite{Toe88a}. Moreover, this
solves part of the double counting problem and recovers the stability
condition (see \cite{Tselyaev_2013}).

\subsection{Selection of most relevant phonons}
\label{sec:select}

In the self-consistent TBA, in which the relations (\ref{sccond}) are
fulfilled, we formally have no free parameters in addition to the
parameters of the energy density functional.  Nevertheless, there is
the question of what number and what kind of phonons should be
included in the $ph\otimes$phonon space of the model. This
question concerns the problem of convergence with respect to enlarging
the $ph\otimes$phonon subspace. So, it is important to select {a
  small amount of} phonons producing the strongest coupling between
the $ph$ and $ph\otimes$phonon configurations.  {Moreover,
  including only sufficiently collective phonons minimizes violation
  of the Pauli principle and reduces} the problem of double counting
connected to the second-order contributions.  {However, the quest
  of a} well-justified and clear criterion of collectivity is
  still matter of debate.

For example, the values of the contributions of the separate $ph$
components (first of all, the main component) of the RPA transition
amplitude $z^n_{12}$ for a positive frequency state,
$\omega^{\vphu}_n>0$, into the norm (\ref{zmz}) can be chosen as
this criterion (see, e.g.,
Refs.~\cite{Bertulani_1999,Litvinova_2007pdr}).  {This is
  quantified by the quantity $\xi^2_n$}
\be
\xi^2_n =
\max_{(ph)} \sum_{m_p,m_h}\bigl( |z^n_{ph}|^2 - |z^n_{hp}|^2 \bigr)\,,
\label{xi2defa  }
\ee
where $m_p$ and $m_h$ are the projections of the total angular
momentum of the particles and holes, $(ph)$ denotes the set of the
remaining single-particle quantum numbers. For the non-collective
phonons the value of $\xi^2_n$ should be close to 1. The maximum value
of $\xi^2_n$ for the collective phonons ($\xi^2_{\mbsu{max}}$)
{lies typically} in the interval 0.5-0.6 \cite{Bertulani_1999}.
This method estimates the spread of the RPA state over the $1p1h$
configurations, {but it} does not introduce any energy cutoff of
the phonon's basis that would be desirable to limit the {phonon
  subspace}.  In addition, this criterion does not give information
about the magnitude of the particle-phonon coupling for the selected
phonons.

Frequently, {also in our earlier work,} the probability criterion is
used in which the phonons are selected according to the values of the
reduced probability of the electric transition $B(EL)$ calculated with
the transition amplitude $|\,{z}^{n} \rangle$ of the phonon (see,
e.g., \cite{Colo_1994,Kamerdzhiev_2004,Litvinova_2007,Tselyaev_2016}).
This means that {only phonons} with
$B(EL)/B(EL)_{\mbsu{max}}>b_{\mbsu{cut}}$ are included in the phonon
basis, where $B(EL)_{\mbsu{max}}$ is the maximal $B(EL)$ for the given
multipolarity $L$, $b_{\mbsu{cut}}$ is the cutoff parameter typically
ranging from $1/10$ to $1/5$. In Ref.~\cite{Colo_1994} this criterion
is formulated in terms of the phonon's contribution to the energy
  weighted sum rule (EWSR).  This integral method of the selection of
the phonons (we will refer to it as the $B$-criterion) is based on the
assumption that the excitation modes having the largest transition
probabilities {are the most collective ones and} should have the
strongest coupling to the single-particle states (see discussion at
the end of this section). However, the connection of the $B$-criterion
to collectivity is, {in fact, not so obvious}.

Here we suggest another method in which the connection to
{collectivity and interaction strength} becomes more explicit.
Let us {introduce the average interaction strength in mode $n$
and average {$ph$} energy as}
\be
\langle\,V\,\rangle^{\vphu}_n =
\langle\,{z}^{n}\,|\,V\,|\,{z}^{n} \rangle\,,
\label{vmn}
\ee
\be
|\,\omega^{(0)}_{n}| =
\langle\,{z}^{n}\,|\,M^{\mbss{RPA}}_{\vphd}
\Omega^{(0)}_{\vphd}|\,{z}^{n} \rangle\,.
\label{omzn}
\ee
In terms of the basis (\ref{spbas}) we have:
\be
|\,\omega^{(0)}_{n}| = \sum_{ph}\,(\ve^{\vphu}_{p}-\ve^{\vphu}_{h})
\bigl(\,|\,z^{n}_{ph}\,|^2 + |\,z^{n}_{hp}\,|^2 \bigr)\,.
\label{omznph}
\ee
From Eqs.~(\ref{rpazes}) and (\ref{zmz}) we obtain
\be
\langle\,{z}^{n}\,|\,M^{\mbss{RPA}}_{\vphd}
\Omega^{\mbss{RPA}}_{\vphd}|\,{z}^{n} \rangle =
|\,\omega^{\vphu}_{n}|\,.
\label{zmomz}
\ee
From Eqs. (\ref{omrpas}), (\ref{vmn}), (\ref{omzn}), (\ref{zmomz}) and from
the property $(M^{\mbss{RPA}}_{\vphd})^2 = 1$ in the $ph$ space
it follows that
\be
\langle\,V\,\rangle^{\vphu}_n =
|\,\omega^{\vphu}_{n}| - |\,\omega^{(0)}_{n}|\,.
\label{vmnf}
\ee As follows from Eq.~(\ref{vmn}), the quantity
$\langle\,V\,\rangle^{\vphu}_n$ {represents} the average value of
the residual interaction in the RPA state $|\,{z}^{n} \rangle$. The
values of $\langle\,V\,\rangle^{\vphu}_n$ can be easily calculated
using Eqs. (\ref{omznph}) and (\ref{vmnf}) if the solutions of the RPA
equation are known.

The new criterion for selection of phonons (positive
frequency: $\omega^{\vphu}_{n} > 0$) is
\be
  |\,v^{\vphu}_n\,| > v_{\mbsu{min}}
  \;,\;
  v^{\vphu}_n =
  \langle\,V\,\rangle^{\vphu}_n / \omega^{\vphu}_{n}
  \;.
\label{vndef}
\ee
This means that only those phonons will be included in the TBA basis
whose dimensionless interaction strength $v^{\vphu}_n$ exceeds the
cutoff value $v_{\mbsu{min}}$. Note that the negative (positive) sign of
$v^{\vphu}_n$ indicates that the residual interaction in the state
$|\,{z}^{n} \rangle$ has on average attractive (repulsive) character.
 We call this selection the $V/E$-criterion in the following.  

There are several arguments to justify the $V/E$-criterion.
  First, from Eqs. (\ref{gndef}) and (\ref{vmn}) we obtain
\be
\langle\,V\,\rangle^{\vphu}_n = \langle\,{z}^{n}\,|\,{g}^{n} \rangle\,.
\label{vmnzg}
\ee
In the macroscopic approach (see, e.g., Refs.~\cite{BM2,Bertsch_1983})
both the transition amplitudes $z^{n}_{12}$ and the amplitudes
$g^{n}_{12}$ are proportional to the {dimensionless deformation
  amplitudes} $\beta^{\vphu}_n$ of the respective vibrational modes.
So, in this approach $|\,v^{\vphu}_n\,| \propto \beta^2_n$.
Therefore, the selection of the phonons with the largest values of
$|\,v^{\vphu}_n\,|$ corresponds to the selection of the low-energy
vibrational modes with the largest deformation {amplitudes having
  thus} the strongest coupling to the single-particle states.

Second, we take the point of view that $|\,v^{\vphu}_n\,|$ is a
measure of collectivity.  Large values make a large (collective)
shift of energy as shown by Eq. (\ref{vmnf}). Small values 
correspond to uncorrelated (non-collective) RPA modes which are 
dominated by what is called a one-loop diagram. 
But just such uncorrelated RPA modes taken as the phonons produce 
the second-order contributions in the response
function of the TBA which should be eliminated to avoid double
counting \cite{Row70aB}.

Third, the $V/E$-criterion seems to be preferable as compared with the
$B$-criterion {because it requires no additional assumptions.
  Note that the $B$-criterion relies on $B(EL)$ values which refer to
  an external multipole operator and it requires additional cutoff
  value $L^{\mbsu{phon}}_{\mbsu{max}}$ and multipolarity whereas the
  $V/E$-criterion is the same for all the multipolarities and
  automatically eliminates all states with too large values of $L$.}

Nevertheless, {it is interesting to establish a} connection
between the $V/E$- and $B$-criteria. {Consider a} residual
interaction $V$ in a separable form as a multipole decomposition (see,
e.g., \cite{Soloviev_1976})
\be
V = \sum_{\alpha,L,M} \kappa^{\alpha}_{L}\,|\,Q^{\alpha}_{LM} \rangle
\langle\,Q^{\alpha}_{LM}\,|\,,
\label{sepint}
\ee
where $Q^{\alpha}_{LM}$ are the multipole operators, index $\alpha$
labels different kinds of these operators (electric, magnetic,
isoscalar, isovector, etc.), $\kappa^{\alpha}_{L}$ are the force
parameters.  If {the same $Q^{\alpha}_{LM}$ is taken as the
multipole operator for the $B(EL)$ value,} the respective reduced
probability in the RPA reads
\be
B_n (\alpha L_n) = \sum_{M_n} |\langle\,{z}^{n}\,|\,Q^{\alpha}_{L_n M_n} \rangle|^2.
\label{redprob}
\ee
From Eqs. (\ref{vmn}), (\ref{sepint}) and (\ref{redprob}) we obtain
\be
\langle\,V\,\rangle^{\vphu}_n =
\sum_{\alpha} \frac{\kappa^{\alpha}_{L_n}}{2L_n + 1}
\,B_n (\alpha L_n)\,.
\label{vsepmn}
\ee
Thus the $B(EL)$ values are indeed strongly related to {the} average
  interaction strength $\langle\,V\,\rangle^{\vphu}_n$.  There are,
  however, different weight factors which impact the criteria. 
The $V/E$-criterion employs a weight
$\kappa^{\alpha}_{L_n}(2L_n + 1)^{-1}\omega_{n}^{-1}$
(if Eq.~(\ref{vsepmn}) is fulfilled and the phonon state selects
a certain value of $\alpha$) while the $B$-criterion
is weighted with $B(EL)_{\mbsu{max}}^{-1}$.
  Different weights create different
  selectivity and we have yet to see how the two criteria compare in
  practice. Realistic residual interactions are not strictly
  separable, but are found often rather close to a sum of separable
  terms \cite{Nes02a}. Thus the above relation maintains some general
  relevance.

\section{Details of calculations}
\label{detailscalculation}

\subsection{Basic Equations}
\label{sec:basicEq}

Our approach is based on the version of the response function
formalism developed within the Green function method (see
\cite{Shlomo_1975,Speth_1977}).  Details are described in
\cite{Tselyaev_2016,Lyutorovich_2015}. 
The basic calculated quantity is the response $R(\omega)$ which is a solution
of the Bethe-Salpeter equation.  In RPA and TBA, it reads
\be
R^{\mbss{RPA}}_{\vphd}(\omega) = R^{(0)}_{\vphd}(\omega)
- R^{(0)}_{\vphd}(\omega)VR^{\mbss{RPA}}_{\vphd}(\omega)\,,
\label{rfrpa}
\ee
\bea
R^{\mbss{TBA}}_{\vphd}(\omega) &=& R^{(0)}_{\vphd}(\omega)
\nonumber\\
&-& R^{(0)}_{\vphd}(\omega)\bigl(\,V + \bar{W}(\omega)\bigr)
R^{\mbss{TBA}}_{\vphd}(\omega)\,,
\label{rftba}
\eea
where
\be
R^{(0)}_{\vphd}(\omega) = -\bigl(\omega - \Omega^{(0)}\bigr)^{-1}
M^{\mbss{RPA}}_{\vphd}
\label{rfuncor}
\ee
is the uncorrelated $ph$ propagator. The
$ph$-interaction $V$ is defined in Eq. (\ref{sccond}) and the induced
interaction $\bar{W}(\omega)$ in Eqs. (\ref{wsub},\ref{wdef}).

Knowledge of the response function allows us to calculate the
distribution of the nuclear transition strength $S(E)$ caused by an
external field which is represented by a single-particle operator
$Q$. It reads
\be
S(E)=-\frac{1}{\pi}\;\mbox{Im}\,\Pi(E+i\Delta)\,,
\label{sfdef1}
\ee
\be
\Pi(\omega) = - \langle\,Q\,|\,R(\omega)\,|\,Q\,\rangle\,,
\label{poldef1}
\ee
where $E$ is an excitation energy, $\Delta$ is a smearing parameter,
and $\Pi(\omega)$ is the (dynamic) polarizability.

\subsection{Details of the numerical treatment}
\label{sec:numer}

The two approaches studied here, RPA and TBA, are realized with
the same numerical representation. The $sp$ energies,
$sp$ wavefunctions, and the residual $ph$ interaction are obtained from
stationary Skyrme-Hartree-Fock (SHF) calculations based on the
  Skyrme energy functional.  All terms of the residual interaction
are treated in the RPA and TBA {fully self-consistently},
according to the formulas given in Ref.~\cite{Lyutorovich_2016}, but
with one exception: the spin-spin terms are omitted. This does not
break self-consistency because the terms are not active in the ground
state of the double-magic nuclei.  The new treatment of the
single-particle continuum \cite{Tselyaev_2016} is used both in the RPA
and TBA calculations. {There,} the RPA and the TBA equations are
solved in a discrete basis that simplifies calculations of matrix
elements of the residual interaction.  However, the uncorrelated {$ph$}
propagator $R^{(0)}_{\vphd}(\omega)$ is constructed with Green
functions {properly taking into account the nucleon continuum}.
Therefore {the uncorrelated as well as the correlated propagators} 
do not contain discrete
poles at positive energies but a smooth cut.

The box sizes were 15~fm for $^{16}$O, $^{40, 48}$Ca and 18~fm for
$^{56}$Ni, $^{132}$Sn, $^{208}$Pb. The maximal angular momentum of the
$sp$ basis was limited to $l^{\mbsu{sp}}_{\mbsu{max}}$ =17 which was
found by several tests to be a sufficiently large value.  We checked
the dependence of the results on the maximum s.p. energy
$\ve^\mathrm{sp}_\mathrm{max}$.  For light ($^{16}$O) and medium mass
nuclei (Ca and Ni) we found saturation at
$\ve^\mathrm{sp}_\mathrm{max}$ = 500 MeV, and {for} heavy nuclei
$^{132}$Sn and $^{208}$Pb at $\ve^\mathrm{sp}_\mathrm{max}$ = 100 MeV.
The maximal $sp$ energies were thus set as:
$\ve^\mathrm{sp}_\mathrm{max}= 500$ MeV for $^{16}$O, $^{40, 48}$Ca,
$^{56}$Ni and $\ve^\mathrm{sp}_\mathrm{max} = 100$ MeV for $^{132}$Sn
and $^{208}$Pb.  The phonon basis is restricted by the cutoff
$v_{\mbsu{min}}$, or {$B(EL)$} respectively, and the dependence on the
cutoff will be studied in section \ref{dependencies}.  The maximal
phonon energy is $E^{\mbsu{phon}}_{\mbsu{max}}$=40 MeV.  The maximal
phonon angular momentum $L^{\mbsu{phon}}_{\mbsu{max}}$ is determined
by these conditions.  The maximal angular momentum of the
quasiparticle-phonon configurations is 
$L^{1p1h\otimes\mbsu{phon}}_{\mbsu{max}}=27$.
In the following, we keep these parameters of representation
($sp$ space, maximum
  phonon energy, and maximum angular momentum) fixed at rather large
  values because we want to concentrate on the trends with the phonon
  cutoff $v_\mathrm{min}$, or $B(EL)$ respectively. The question of
  convergence with these parameters of representation and possible
  saving at this site will spared for a forthcoming publication.

Each resonance positions were characterized by the energy centroid
defined as the ratio $E_0=m_1/m_0$ of the first and zeroth energy
moments of the corresponding strength $S(E)$. For the GDR (here we
considered the photo absorption cross section) as well as for the GMR
and GQR in $^{132}$Sn, $^{208}$Pb (here we considered the fraction of
the EWSR), the centroids were calculated in the energy windows $E_0
\pm 2\delta $ where $\delta$ was the spectral dispersion.  To avoid
too small energy windows, we used the constraint $\delta >
\delta_{\text{min}}$ where $\delta_{\text{min}}=2.5$ MeV for the GDR
in $^{16}$O, \ 2 MeV for the GDR in $^{40,48}$Ca, $^{132}$Sn,
$^{208}$Pb and for all resonances in $^{56}$Ni, \ 0.5 MeV for the GMR
and GQR in $^{132}$Sn, $^{208}$Pb.  The width $\Gamma$ and dispersion
for these resonances were defined as
\be 
  \Gamma = 2\delta \sqrt{2\ln 2}, 
  \quad 
  \delta = \frac{\int (E-E_0)^2\, S(E)\, dE}{m_0}
\ee 
These $E_0$ and $\Gamma$ are
approximate values of the Lorentzian parameters.

The isoscalar strengths in light nuclei are distributed in large
energy ranges and have a complex structure therefore for these
strengths we used the large windows:\\ $11<E<40$ MeV for GMR and GQR
in $^{16}$O,\\ $10 < E < 30$ MeV for GMR in $^{40, 48}$Ca,\\ and $10 <
E < 25$ MeV for GQR in $^{40, 48}$Ca.  The peak position of the
low-lying dipole strength in $^{132}$Sn was determined as the energy
with the maximum cross section.  The resonance widths $\Gamma$ depend
slightly on the smearing parameter $\Delta$. In all the RPA and TBA
calculations, we used in Eq.~(\ref{sfdef1}) $\Delta = 400$ keV. It is
questionable to represent the broad and strongly fragmented spectral
distribution in $^{16}$O by only two numbers (peak energy, width). But
it suffices for the purpose of comparison because we handle the
experimental data the same way.

\subsection{Choice of Skyrme Parametrization}
\label{sec:skyrme}

From the variety of self-consistent nuclear mean-field models
\cite{Bender_2003}, we consider here the Skyrme-Hartree-Fock (SHF)
functional, for a detailed description see
\cite{Bender_2003,Stone_2007,Erler_2011}. Its essential features are:
The functional depends on a couple of local densities and currents
(density $\rho$, gradient of density $\nabla\rho$, kinetic-energy
density $\tau$, spin-orbit density $\vec{J}$, current $\vec{j}$, spin
density $\vec{\sigma}$, kinetic spin-density $T$). All densities and
currents exist twofold, for isospin zero and isospin one. It consists
of quadratic combinations of these local quantities, corresponding to
pairwise contact interactions.  In principle, all parameters in front
of these contact terms could be density dependent. In practice,
density dependence is considered only for the term $\propto\rho^2$
(both isospins).  Pairing is incorporated by adding a separate pairing
functional.  The typically 13--14 model parameters are considered as
being universal parameters applying throughout the whole nuclear
landscape and bulk matter.  They are determined by a fit to a large
body of experimental data of the nuclear ground state (binding
energies, radii, spin-orbit splittings, pairing gaps). For recent
examples see \cite{Goriely_2002,Kluepfel_2009,Kortelainen_2010}.

Nuclei span a large range in mass number, but a small one in
neutron-proton difference. Their ground states are stationary states.
This means that the ground state fits determine predominantly
isoscalar static properties and leave some leeway in other respects.
Thus there exist many Skyrme parametrizations which perform comparably
well in ground state properties but differ in the less well determined
aspects amongst them many response properties, equivalent to nuclear
matter properties (NMP).  As a consequence, a study using Skyrme
forces should employ a couple of different parametrizations to explore
the possible variety of predictions. To quantify the variation, we
consider the key response properties of the forces in terms of NMP,
i.e. equilibrium properties of symmetric nuclear matter, namely
incompressibility $K$ (isoscalar static), effective mass $m^*/m$
(isoscalar dynamic), symmetry energy $a_\mathrm{sym}$ (isovector
static), TRK sum rule enhancement $\kappa_\mathrm{TRK}$ (isovector
dynamic). These four NMP have a one-to-one relation to nuclear giant
resonances in $^{208}$Pb \cite{Kluepfel_2009}: $K$ to GMR, $m^*/m$ to
GQR, $\kappa_\mathrm{TRK}$ to GDR, and $a_\mathrm{sym}$ to dipole
polarizability \cite{Nazarewicz_2013}.  In order to allow well defined
explorations, the survey \cite{Kluepfel_2009} provides a series of
Skyrme parametrizations with systematically varied NMP.  The present
survey aims at exploring the effect of phonon coupling on excitation
properties. Here, the response properties are crucial and we take a
minimal subset of these systematically varied parametrizations to
discriminate robust features from changing ones.
\begin{table}
\centering
\begin{tabular}{l|cccc}
  & $K$ [MeV]& $m^*/m$ & $a_\mathrm{sym}$ [MeV]& $\kappa_\mathrm{TRK}$ \\
\hline
SV-bas &  234 &  0.90  &  30  &  0.4 \\
SV-mas07 & 234 &  0.70  &  30  &  0.4 \\
SV-m64k6 & 241 & 0.64  & 27 & 0.6\\
\hline
\end{tabular}
\caption{\label{tab:NMP} Nuclear matter properties (NMP) for the three
  Skyrme forces used here: incompressibility $K$, isoscalar effective
  mass $m^*/m$, symmetry energy $a_\mathrm{sym}$, Thomas-Reiche-Kuhn
  sum rule enhancement $\kappa_\mathrm{TRK}$. Parametrizations SV-bas,
  and SV-mas07 \cite{Kluepfel_2009}, SV-m64k6 from
  \cite{Lyutorovich_2012}.}
\end{table}
Table \ref{tab:NMP} lists the chosen parametrizations and their
NMP. SV-bas is the base point of the variation of forces. Its NMP are
chosen such that dipole polarizability and the three most important
giant resonances (GMR, GDR, and GQR) in $^{208}$Pb are well reproduced
by RPA calculations. 
SV-mas07 varies the effective mass while keeping the other NMP fixed.
SV-m64k6 was developed in \cite{Lyutorovich_2012} with the goal
to describe, within TBA, simultaneously the GDR in $^{16}$O and
$^{208}$Pb. This required to push up the RPA peak energy in $^{208}$Pb
which was achieved by low $a_\mathrm{sym}$ in combination with high
$\kappa_\mathrm{TRK}$. To avoid unphysical spectral bunching for the
GDR, a low $m^*/m$ was used.

\section{Dependencies}
\label{dependencies}

In this section, we investigate the dependence of mean energy and
width of giant resonances on
the cutoff parameter for the phonon space as was introduced in
Section \ref{newcriterion}.
Recall, that the crucial ingredient in the phonon-coupling model
  is the number of active phonons. As already discussed in Section
  \ref{newcriterion}, we propose as selector the parameter
  $v_{\mbsu{min}}$ which is connected with the collectivity of the
  phonons. Large $v_{\mbsu{min}}$ exclude automatically phonons which
are dominated by one or two $ph$ components only.

\subsection { Comparing $B$-criterion with $V/E$-criterion}

\begin{figure}
\centerline{\includegraphics[width=0.8\linewidth]{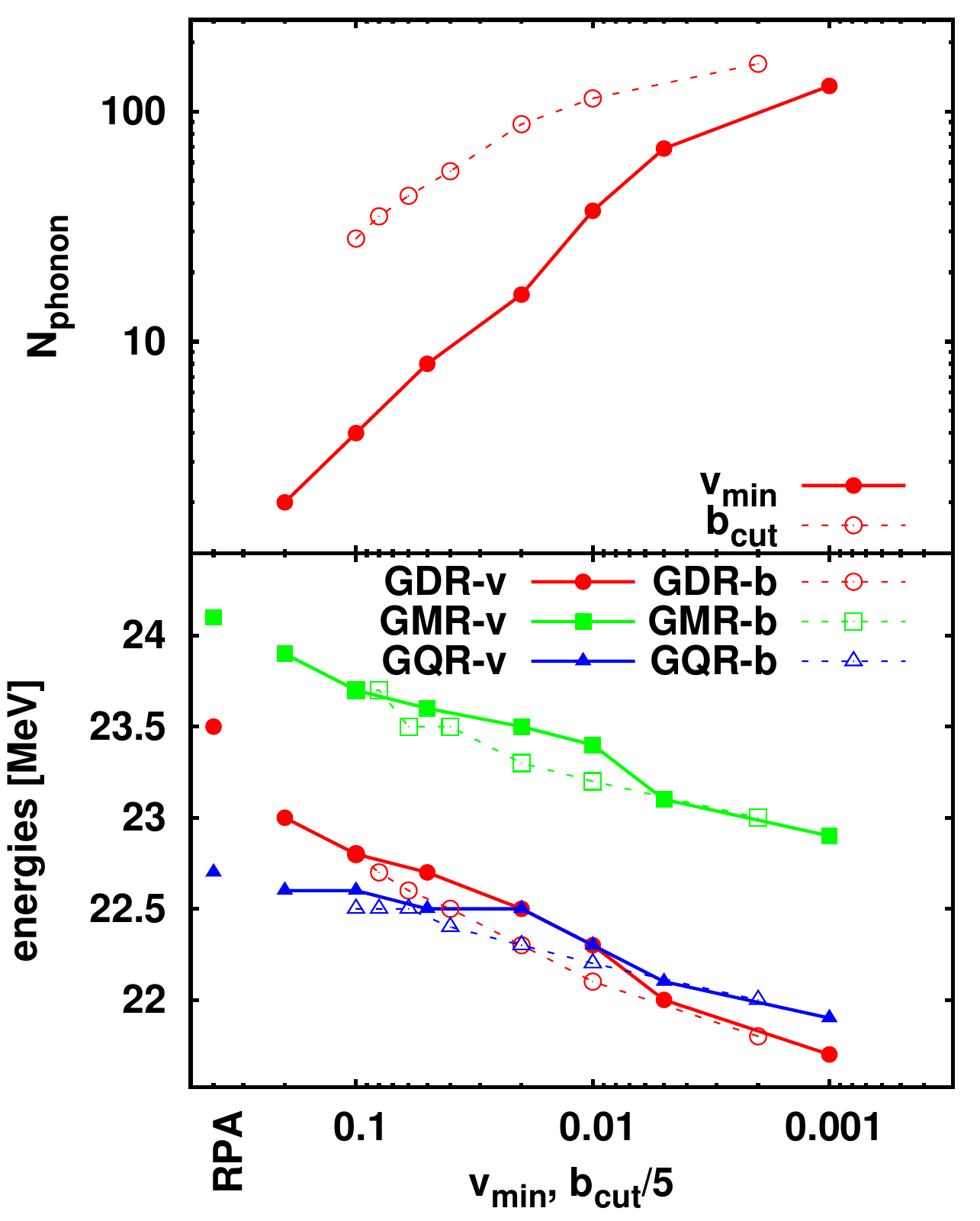}}
\caption{\label{fig:compare-vminbcut-16O} Trends of TBA results as
  function of cutoff criterion $v_\mathrm{min}$ and $b_\mathrm{cut}$
  in comparison, computed for $^{16}$O with the Skyrme parametrization
  SV-m64k6.  Upper panel: Number of phonon states. Lower panel: Mean
  energies of the giant resonances where the appendix ``-v'' indicates
  computation with $v_\mathrm{min}$ cutoff and ``-b'' $b_\mathrm{cut}$
  cutoff. On the left side of each Figure are the corresponding RPA
  values displayed, as indicated.}
\end{figure}

Fig. \ref{fig:compare-vminbcut-16O} compares mean energies of GR
(lower panel) and number of active phonons (upper panel) as function
of the cutoff parameters $v_\mathrm{min}$ and $b_\mathrm{cut}/5$.  As
one can see, the two criteria give very similar results for the GR
energies (lower panel) when scaling $b_\mathrm{cut}$ by factor $1/5$.
However, the number of phonons (upper panel) is much different.  The
$V/E$-criterion achieves the same GR peak energies with substantially
less phonons. This indicates that the $V/E$-criterion is more
efficient in selecting the relevant phonons.

\begin{figure}
\centerline{\includegraphics[width=\linewidth]{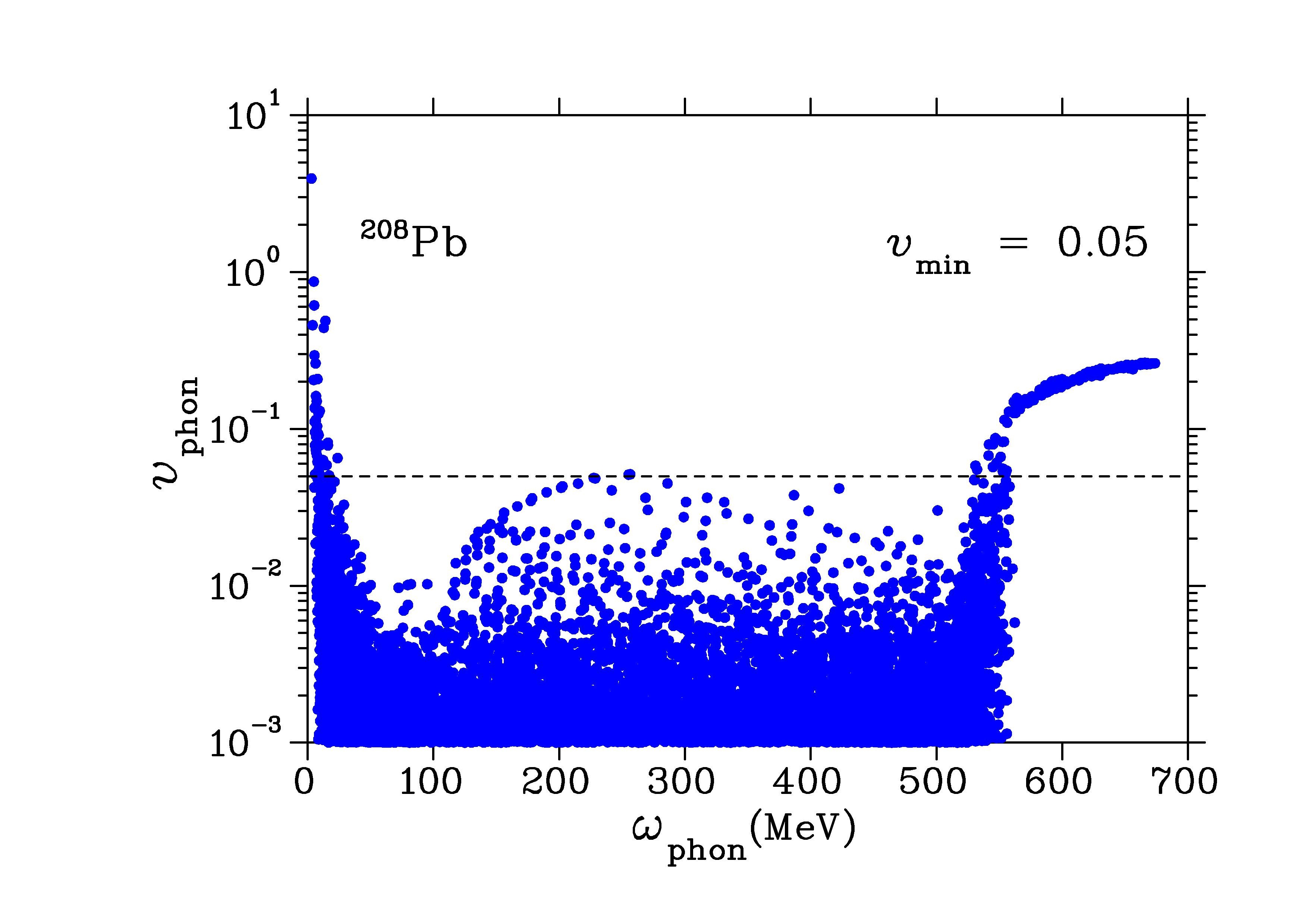}}
\caption{\label{fig:pb208e500-vmin1}
    (Color online) Scatter plot of the distribution of scaled phonon strength
    $v_{\mbsu{phon}}=|v_n|$ defined by Eqs. (\ref{vmn}) and (\ref{vndef})
    (blue circles) as function of phonon energy
    $\omega_{\mbsu{phon}}$ for $^{208}$Pb (see text for more details).
    The dashed horizontal line indicates a natural
    cutoff $v_{\mbsu{min}} =0.05$ for the selection of phonons in the TBA.}
\end{figure}

\subsection {Dependence on the cutoff $v_{\mbsu{min}}$}
\label{dependencies2}

The distribution of relative strength of the phonon's states $|v_n|$
defined by Eq. (\ref{vndef}) in $^{208}$Pb is shown in
Fig.~\ref{fig:pb208e500-vmin1}. The calculations were performed in the
discrete self-consistent RPA for the Skyrme parametrization SV-m64k6
\cite{Lyutorovich_2012} with $\ve^\mathrm{sp}_\mathrm{max}$ = 500
MeV. For the most collective low-lying vibrational states in
$^{208}$Pb (first $2^+$, $3^-$, $4^+$, $5^-$, and $6^+$ levels), $v_n$
takes the values in the interval from $-4.0$ for the $3^-_1$ state up
to $-0.3$ for the $6^+_1$ state.  The distribution in
Fig.~\ref{fig:pb208e500-vmin1} representing $V/E$-criterion shows a
clear distinction between collective phonons, which stick out well
above background indicated by the dashed horizontal line, and the
non-collective ones below this line.  This suggests that a cutoff at
$v_{\mbsu{min}} =0.05$ is an optimal choice.

The increase of collective states above $\omega_{\mbsu{phon}}=50$ MeV
is an artifact of the Skyrme force. This does not appear if we perform
calculations with a Migdal $ph$-interaction which possesses no
momentum dependence. We have produced similar scatter plots for other
nuclei in this survey and find the same pattern suggesting the
practically the same optimal cutoff. To corroborate this choice, we
will investigate in the following the dependence of results on
resonance energies and width on the choice of cutoff.

\begin{figure}
\centerline{\includegraphics[width=0.85\linewidth]{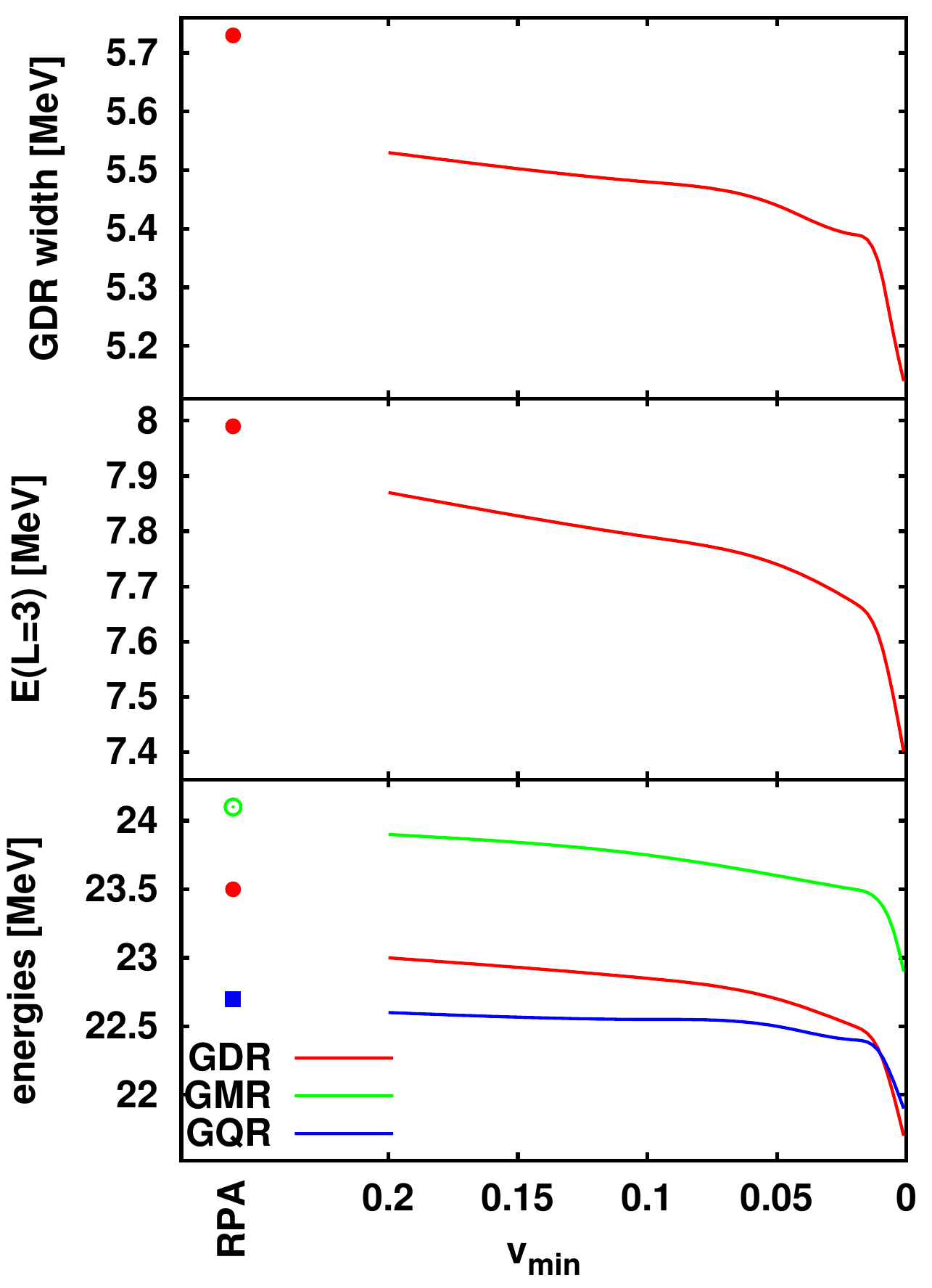}}
\caption{\label{fig:dependences_16O} Mean energies of GR
  resonances (lower panel), excitation energy of the low-lying $3^-$
  resonance (middle panel), and width of GDR (upper panel) for the
  case of $^{16}$O computed with the force SV-m64k4
The results are given as a function of the cutoff criterion
  $v_\mathrm{min}$.  On the left side of each Figure are the
  corresponding RPA values displayed, as indicated  on the
  abscissa.}
\end{figure}

\begin{figure}
\centerline{\includegraphics[width=0.85\linewidth]{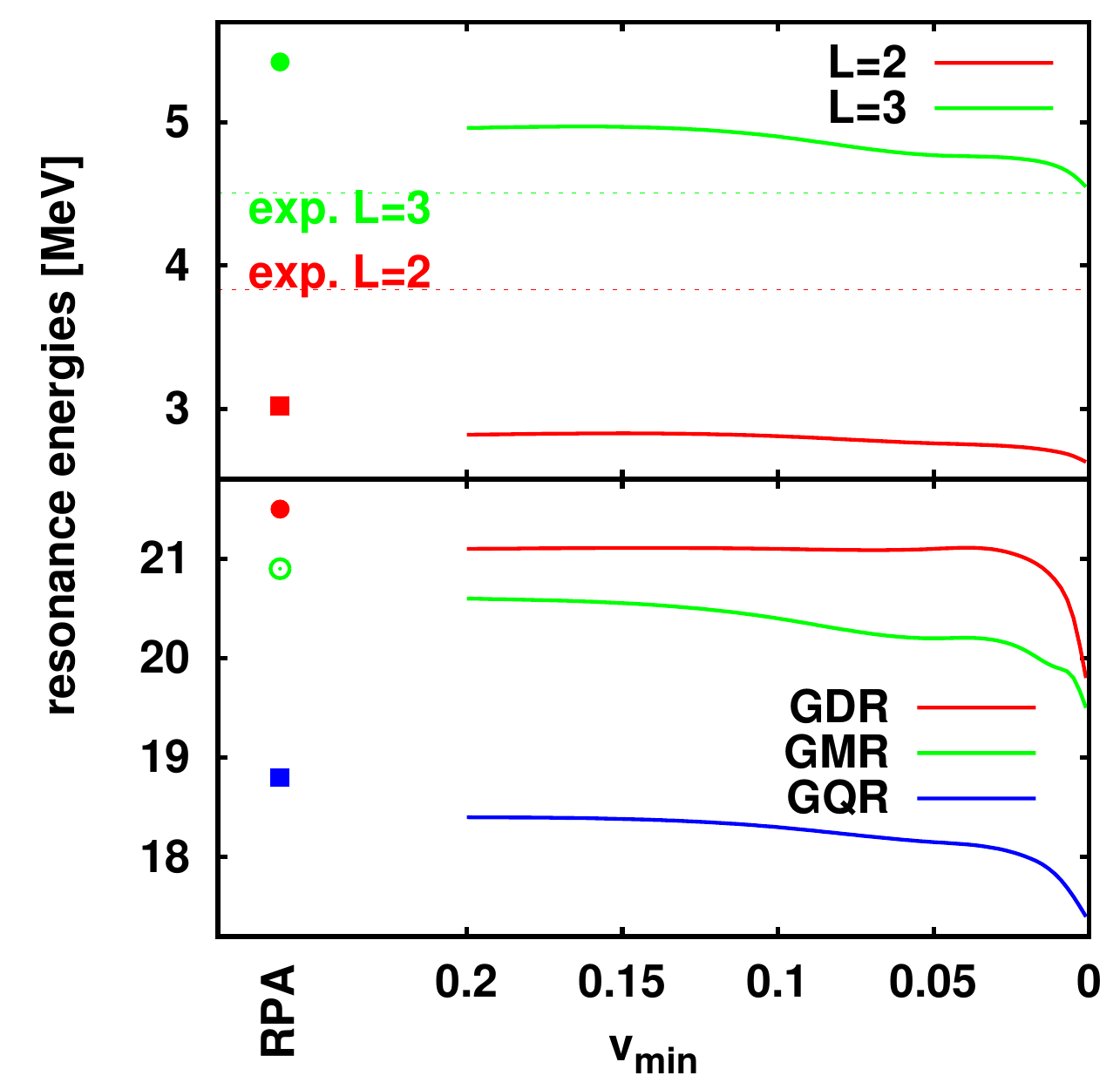}}
\caption{\label{fig:dependences_48Ca} Mean energies of GR resonances
  (lower panel) and excitation energy of the low-lying $2^+$ and $3^-$
  resonances (upper panel) for the case of $^{48}$Ca computed with the
  force SV-m64k4 The results are given as a function of the cutoff
  criterion $v_\mathrm{min}$.  On the utmost left side of each Figure
  are the corresponding RPA values displayed, as indicated on the
  abscissa.}
\end{figure}

\begin{figure}
\centerline{\includegraphics[width=0.85\linewidth]{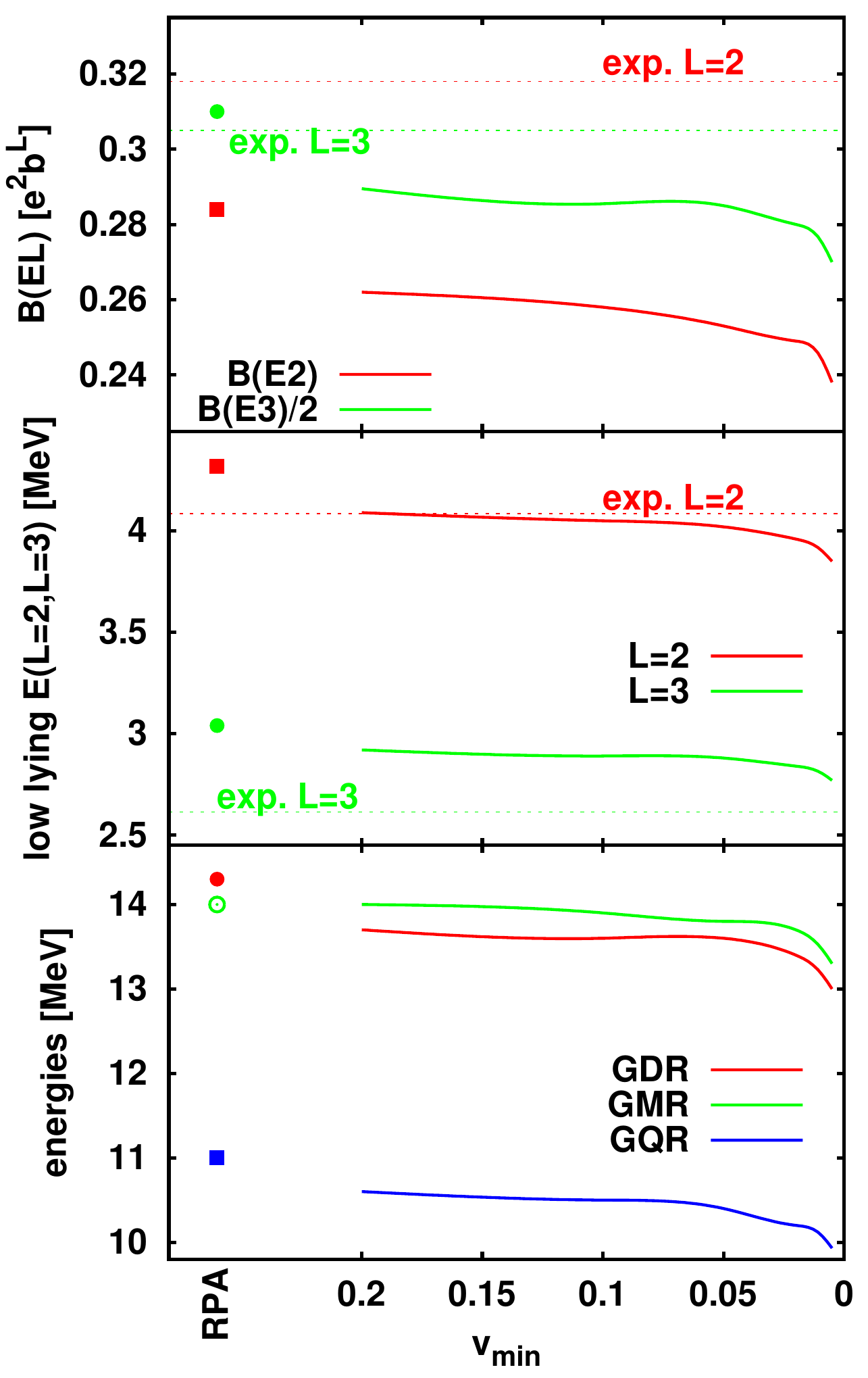}}
\caption{\label{fig:dependences_208Pb_SVbas} Mean energies of the
  giant resonances (lower), excitation energy of the low-lying $3^-$
  state (middle), and their $B(EL)\!\uparrow$ values (upper) for $^{208}$Pb
  computed with the force SV-bas. The results are given as a function
  of the cutoff criterion $v_\mathrm{min}$.  On the left side
  of each Figure are the corresponding RPA values displayed as
  indicated on the abscissa.  }
\end{figure}

Figs. \ref{fig:dependences_16O}--\ref{fig:dependences_208Pb_SVbas}
show the dependence of GR properties and low-lying collective states
in $^{16}$O, $^{48}$Ca, and $^{208}$Pb on the cutoff parameter
$v_{\mbsu{min}}$. The most sensitive dependence on $v_{\mbsu{min}}$ is
seen in the light nucleus $^{16}$O. But even in this case the
variation of these key quantities between the $v_{\mbsu{min}}$ = 0.2
and 0.05 (the value which we finally use) are not very large: The
energies of GR are stable and the variation of the width of the GDR as
well as the excitation energy of the $3^-$ state is still
moderate. Much less dependence on $v_{\mbsu{min}}$ is seen with
increasing mass number $A$. In all cases, we see steep changes
starting sooner or later below $v_{\mbsu{min}}$ = 0.05. Thus we
encounter a plateau of robust results within which we can chose
pertinent $v_{\mbsu{min}}$. We finally take the lower end of the
plateau which complies nicely with the optimum value suggested in the
scatter plot in Fig.~\ref{fig:pb208e500-vmin1}.

\section{Giant Multipole Resonances}
\label{GR}

RPA and TBA were designed to describe nuclear excitation spectra in
the realm of giant multipole resonances \cite{Spe91b} and so GR are to
be the first test case for new developments. We will compare in this
section RPA with TBA results for the three most prominent modes, GDR,
GMR, and GQR. In order to demonstrate the influence of the underlying
energy functional, we will use three different Skyrme parametrizations
as explained in section \ref{sec:skyrme}. Before going on, {let} us
briefly recall basic properties of GR.  The heavier the nucleus the
more concentrated the resonance spectra such that isoscalar GMR and
GQR and isovector GDR display one prominent and rather narrow
resonance peak. Spectral fragmentation takes over for GMR and GQR
towards light nuclei whereas the GDR still remains a compact, though
fragmented, resonance. Finally in $^{16}$O we observe for GMR and GQR
a multi-peak structure, which is distributed over a large energy
band. As mentioned before to define here a \emph{mean energy} is
somewhat questionable. Nevertheless we can evaluate experimental mean
energies the same way which can then be compared with our results.

\begin{figure}
\centerline{\includegraphics[width=\linewidth]{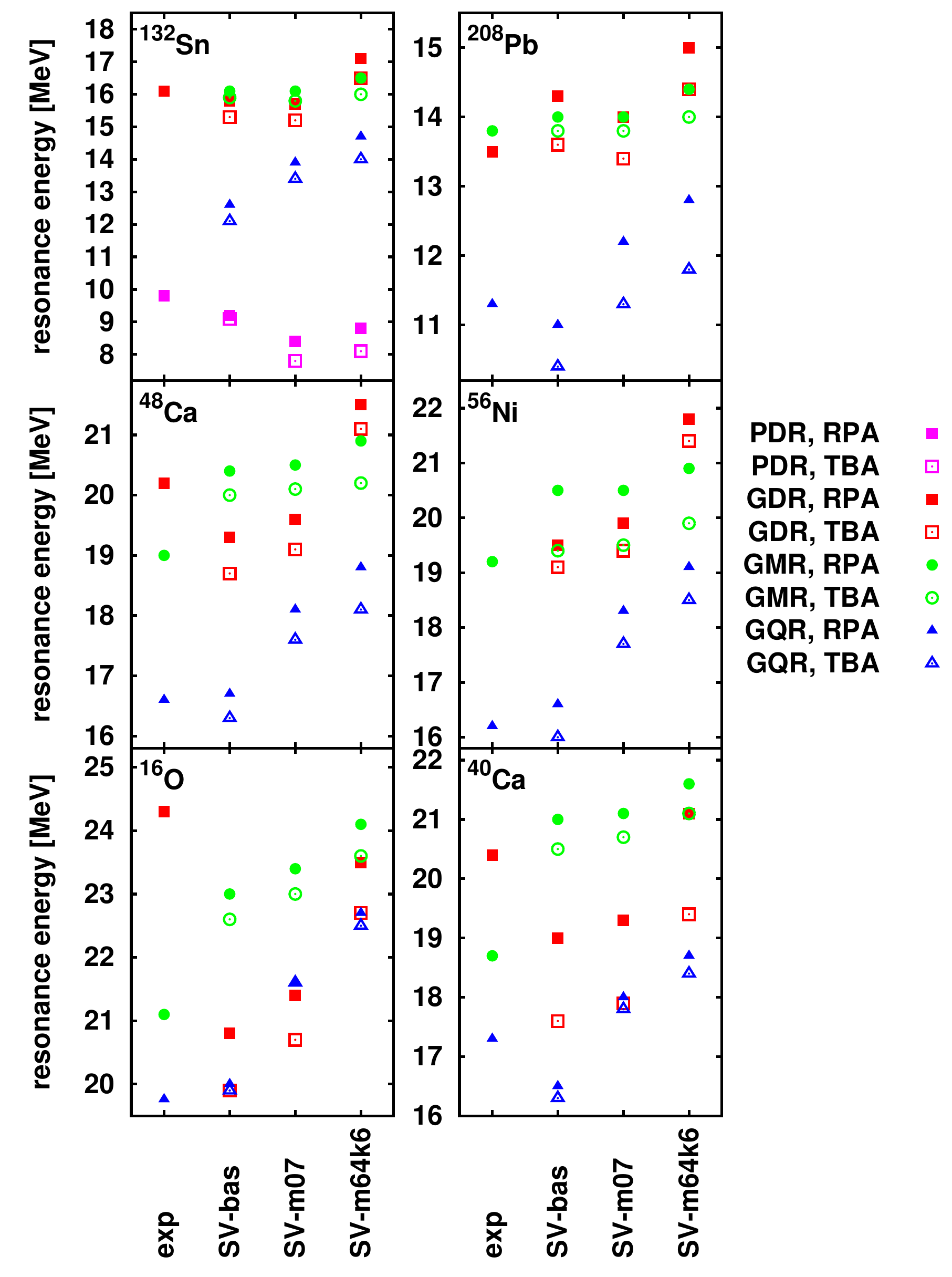}}
\caption{\label{fig:Gr-forces} Mean energies of resonances for three
  Skyrme parametrizations and six doubly magic nuclei. The
    experimental data for GMR and GQR in $^{16}$O were taken from
    Ref.~\cite{Lui_2001} and for GDR in $^{16}$O from
    \cite{Ishkhanov_2002}, in $^{40}$Ca and $^{48}$Ca for GDR from
    \cite{Anders_2013} and GMR/GQR from \cite{Erokhova_2003}, in
    $^{56}$Ni for GMR from \cite{Bagchi_2015,Monrozeau_2008} and for
    GQR from \cite{Monrozeau_2008}, in $^{132}$Sn for GDR from
    \cite{Adrich_2005}, in $^{208}$Pb for GDR from \cite{Belyaev_1995}
    and for GMR/GQR from \cite{Youngblood_2004}.}
\end{figure}
 
Fig.~\ref{fig:Gr-forces} shows the RPA and TBA results on GR
  together with experimental data for the selected nuclei and
  parametrizations. Although our main emphasis lies on the changes
  induced by phonon coupling in the step from RPA to TBA, let us
  briefly comment on the trends with parametrization.  The heavy
  nucleus $^{208}$Pb being closest to bulk displays a nearly
  one-to-one correspondence between NMP and GR peak energies
  \cite{Kluepfel_2009}, namely between GMR and incompressibility, GQR
  and effective mass, GDR and TRK sum-rule enhancement, and dipole
  polarizability and asymmetry energy. We see this in the upper right
  panel of Fig. \ref{fig:Gr-forces}: our set of three parametrizations
  varies $m^*/m$ and accordingly most changing is the GQR, the set
  SV-m64k6 varies additionally $\kappa_\mathrm{TRK}$ and the GDR peak
  comes visibly higher. Smaller nuclei gather increasingly surface
  effects which, in turn, mixes the dependencies. For example, changing
  $m^*/m$ affects all other modes too. Unfortunately, we have to
  realize that the trend with system size $A$ is not reproduced by only
  one of the parametrizations. Take the example SV-bas. It is tuned to
  reproduce GR in $^{208}$Pb but all three GR show large deviations
  for the lightest nuclei in the sample. SV-m64k6 was tuned to cover
  the GDR in $^{16}$O as well as in $^{208}$Pb \cite{Lyutorovich_2012}
  and it does so, but it fails still for GMR and GQR. The problem
  remains yet unsolved that the present form of the Skyrme functional
  cannot yet cover GR in all nuclei \cite{Erl10a}. We will not solve
  it in this paper. The aim is to check the impact of phonon coupling
  on the trends which is also an important ingredient in further
  improving the functional.  

\begin{figure}
\centerline{\includegraphics[width=0.9\linewidth]{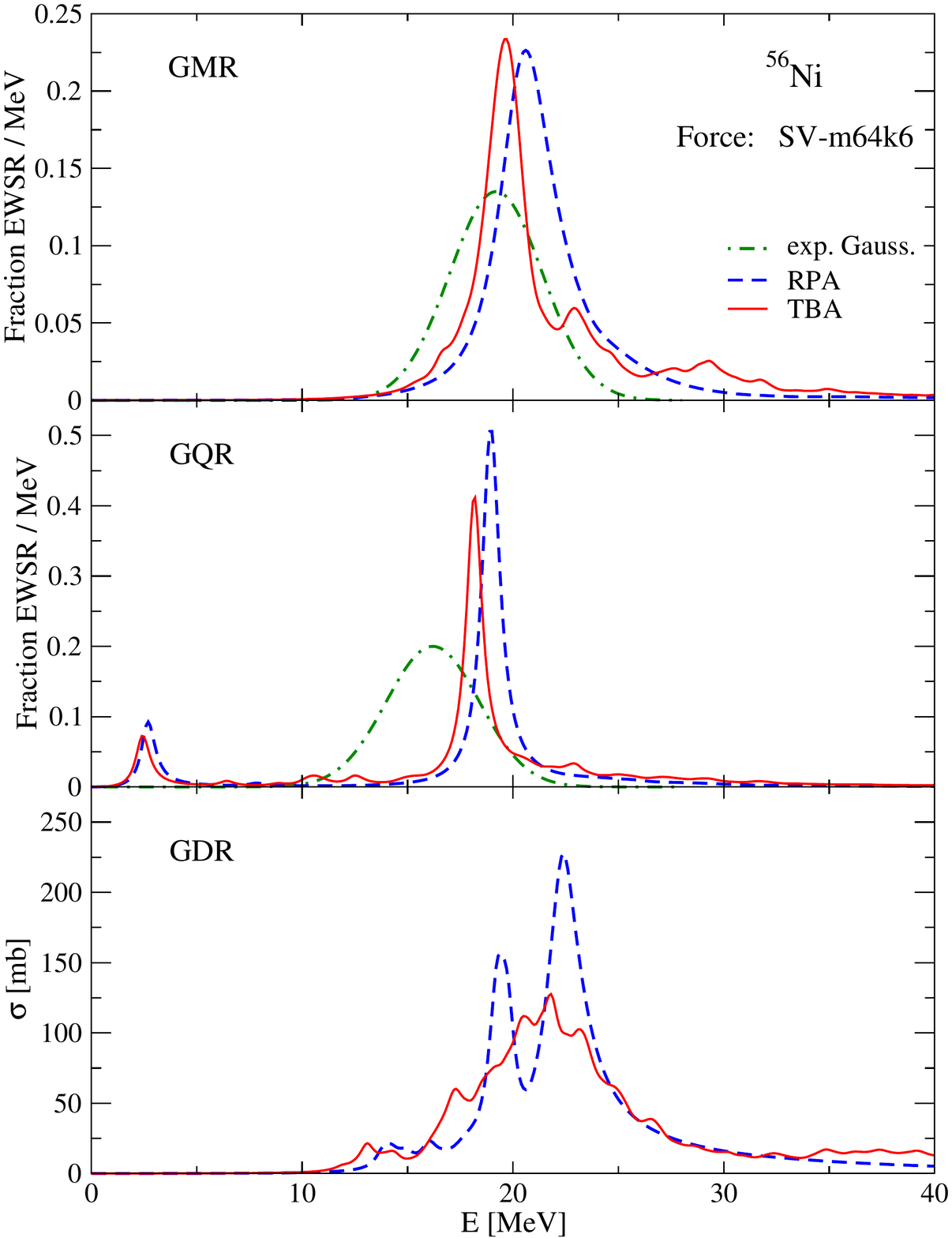}}
\centerline{\includegraphics[width=0.9\linewidth]{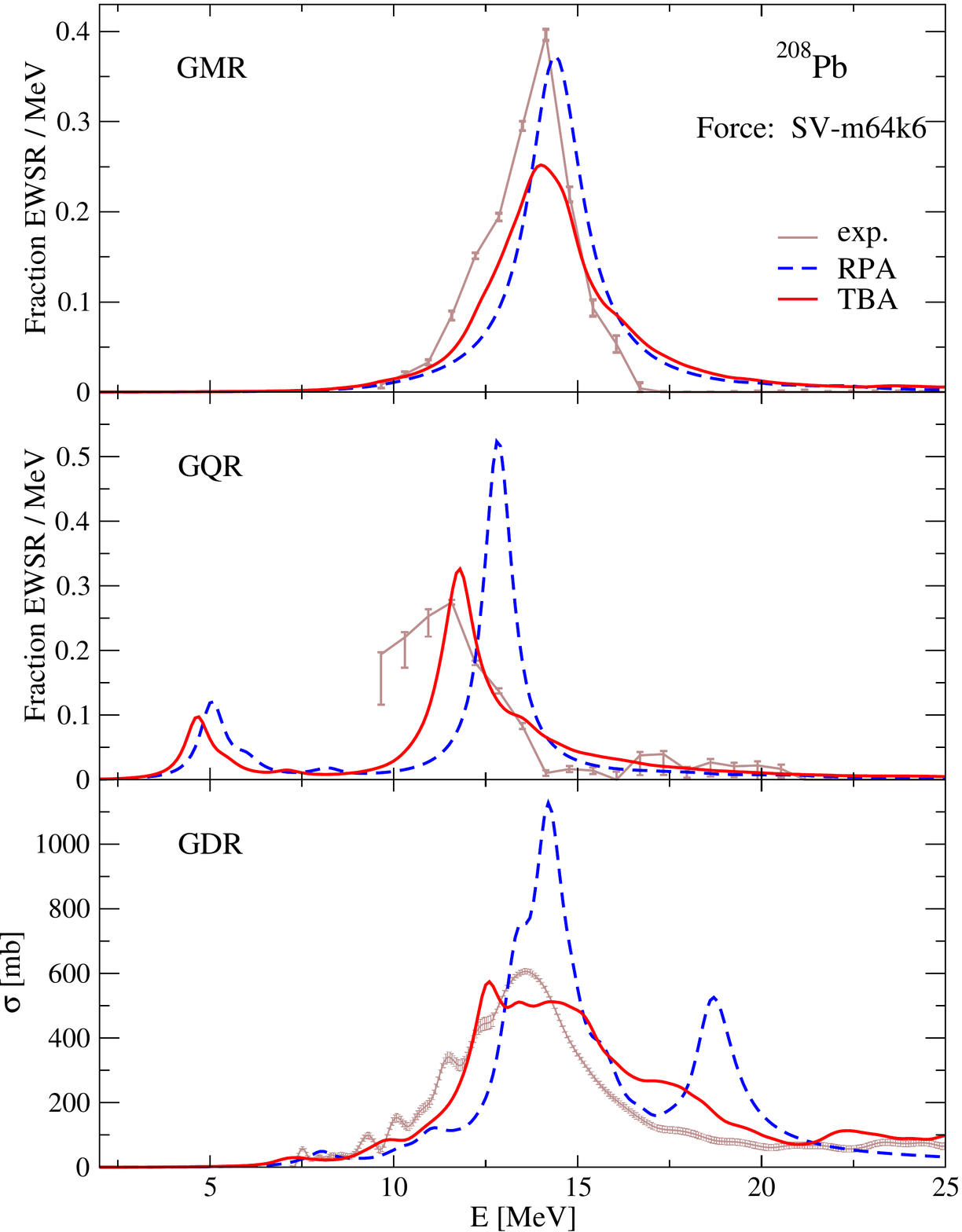}}
\caption{\label{56Ni_strengths} Strength distributions of giant
  resonances in $^{56}$Ni and $^{208}$Pb. Compared are RPA and TBA
  results computed using the Skyrme parametrization SV-m64k6.  The
  blue dashed and full red lines represent RPA and TBA results,
  respectively.  Experimental data (dashed-dotted lines in the upper
  block) are for $^{56}$Ni Gaussians obtained from experimental values
  of $E_0$ and $\Gamma$ \cite{Monrozeau_2008} and (lines with error
  bars in the lower block) for $^{208}$Pb from \cite{Youngblood_2004}
  for GMR and GQR and from \cite{Belyaev_1995} for GDR.  }
\end{figure}

  Now let us look at effect of phonon coupling in
  Fig. \ref{fig:Gr-forces}. The step from RPA to TBA shifts the peak
  energies to lower energies. The prominent result of this summary
  view is that this down-shift is rather small and constant for all
  forces and nuclei. It varies only in a small band of 0.5--1.5 MeV.
  The major effect of phonon coupling remains the smoothing of the
  spectral distributions towards realistic profiles as seen in Fig.
  \ref{pb208} and {Figs. \ref{56Ni_strengths}}.
   The effect on peak energies is much smaller than the
  variations with parametrizations. Thus the solution for the yet
  unresolved trends with $A$ has to come from improved
  functionals. However, the fine-tuning of parametrization should
  stay aware of the small down-shift.  

The structure of the cross sections of the GR resonances are
qualitatively changed by the TBA compared to the RPA. This can not be
seen by comparing the widths of the resonances but one has to look at
the detailed cross sections, or strength distributions respectively.

Fig.~\ref{56Ni_strengths} show
detailed strength distributions for the three GR in $^{56}$Ni and
$^{208}$Pb. It demonstrates the impact of phonon coupling in
detail. The down-shift of peak energy is, though small, well
visible. We see also that the total width is not so much affected.
It is dominated by spectral fragmentation due to 
the different energies of the various $ph$ components. 
This effect, the Landau damping, is already present in RPA and 
in general larger than the escape widths due to the coupling to the continuum. 
TBA adds what is called spreading width or collisional width
\cite{Bertsch_1983} which has moderate impact on the total width, but
shapes significantly the profile of the spectral distribution 
by reducing, in most cases considerably, the height of the main peak.
This feature that phonon coupling works predominantly on the detailed
profile is new as compared to earlier publications. It is due to the
subtraction scheme in Eq.~(\ref{wsub}) which eliminates double counting of
static corrections.

\section{Low-lying collective states}
\label{lowlying}

\subsection{General considerations}

In all nuclei reported here exists a low lying collective
$3^-$~state. As the parity changes from shell to shell one obtains
only low-lying $(ph)$ pairs with negative parity. From $^{48}$Ca on
the spin-orbit splitting is so large that the lower partner of
the first unoccupied shell is shifted beneath the Fermi edge
in the next lower shell. The coupling of this hole state with the
states of the same shell just above the Fermi edge gives rise to
low-lying $ph$ pairs with positive parity. For this reasons, in heavy
nuclei, the gap between the $ph$ states and the spin-orbit splitting
are crucial for the energies of the low-lying states having
negative or positive parity.
The energy gaps between the particle and hole spectra in $^{208}$Pb shown 
in Table II are the differences $1h_{9/2}$ - $3s_{1/2}$ for protons
and $2g_{9/2}$ - $3p_{1/2}$ for neutrons. 
 
\begin{table} 
\begin{ruledtabular}
\begin{tabular}{ccccc}
               & $\Lambda_P$
                       & $\Lambda_N$
                               &$\Delta_P$
                                        & $\Delta_N$ \\
\hline                                                 
               &       &       &        &        \\
$^{208}$Pb     &       &       &        &        \\
SV-bas         & 5.22  & 6.83  &  4.09  &  3.86  \\ 
SV-m64k6       & 5.80  & 7.87  &  4.35  &  5.11  \\
Exp.           & 5.57  & 5.84  &  4.21  &  3.43  \\
               &       &       &        &        \\
\end{tabular}
\end{ruledtabular}
\caption{\label{Spin-Orbit(1)}
Comparison of the energy differences between the spin-orbit($\Lambda$) 
proton partners $1h_{9/2}-1h_{11/2}$ and neutron partners 
$1i_{11/2}-1i_{13/2}$ and energy gaps ($\Delta$) between the particle 
and hole spectra with the experimental data for $^{208}$Pb.
All the values are given in MeV.}
\end{table}
\begin{table}
\begin{ruledtabular}
\begin{tabular}{ccccc}
               & $\Lambda_P$ \footnotemark[1]
                       & $\Lambda_N$ \footnotemark[1]
                               &$\Delta_P$
                                        & $\Delta_N$ \\
\hline                                                 
               &       &       &        &          \\
$^{40}$Ca      &       &       &        &          \\
SV-bas         & 7.06  & 7.45  & 4.99   & 5.10     \\ 
Exp.           & 5.69  & 5.71  & 7.24   & 7.28     \\
               &       &       &        &          \\
$^{48}$Ca      &       &       &        &          \\
SV-bas         & 7.48  & 7.36  & 5.42   & 2.93     \\ 
Exp.           & 5.08  & 8.75  & 6.18   & 4.80     \\
               &       &       &        &         \\
$^{56}$Ni      &       &       &        &          \\
SV-bas         & 7.14  & 7.33  & 4.11   & 4.15    \\ 
Exp.           & 7.45  & 7.17  & 6.42   & 6.40     \\
               &       &       &        &         \\
$^{132}$Sn     &       &       &        &          \\
SV-bas         & 5.59  & 7.42  & 5.59   &  4.68   \\ 
Exp.           & 6.13  & 6.75  & 6.13   &  4.94    \\
               &       &       &        &         \\
\end{tabular}
\end{ruledtabular}
\footnotetext[1]{For $^{132}$Sn, $\Lambda_P$ and $\Lambda_N$
are given for the shells 1$g$ and 1$h$, respectively. 
For all other nuclei, $\Lambda_P$ and $\Lambda_N$ are given for 1$f$.}
\caption{\label{Spin-Orbit(2)}
Comparison of the spin-orbit splitting ($\Lambda$) between the proton 
and neutron partners and the energy gaps($\Delta$) between the particle 
and hole spectra with the experimental data for the double magic 
nuclei $^{40}$Ca,$^{48}$Ca,$^{56}$Ni and $^{132}$Sn. 
All values are given in MeV.}
\end{table}
As we can see from Table~\ref{Spin-Orbit(1)} in the case
of $^{208}$Pb the experimental gaps and the spin-orbit splittings are
nicely reproduced by the SV-bas parametrization whereas the results
from the SV-m64k6 especially for the neutron data deviate strongly
from the experimental values. Therefore we expect that in this
case the results for the low-lying 3$^-$~and 2$^+$~states derived with
the SV-bas parameters agree much better with the data than the ones
obtained with SV-m64k6. In Table~\ref{Spin-Orbit(2)}, we show
spin-orbit splitting and the energy gaps for
$^{40}$Ca,~$^{48}$Ca,~$^{56}$Ni and $^{132}$Sn. Here we compared only
the theoretical results for SV-bas with the data. For the spin-orbit
splitting the agreement between theory and experimental data is good,
with the exception of the proton 1f pair in $^{48}$Ca. Also the energy
gaps between the particle and hole spectra are well reproduced. Here
we have to bear in mind that the SV-bas parameter set was not adjusted
to any of these quantities but to the usual nuclear matter properties
and the monopole, quadrupole and dipole giant resonances~\cite{SV-bas}.

\subsection {Effective mass, spin-orbit splitting and phonons}

\begin{figure}
\centerline{\includegraphics[width=\linewidth]{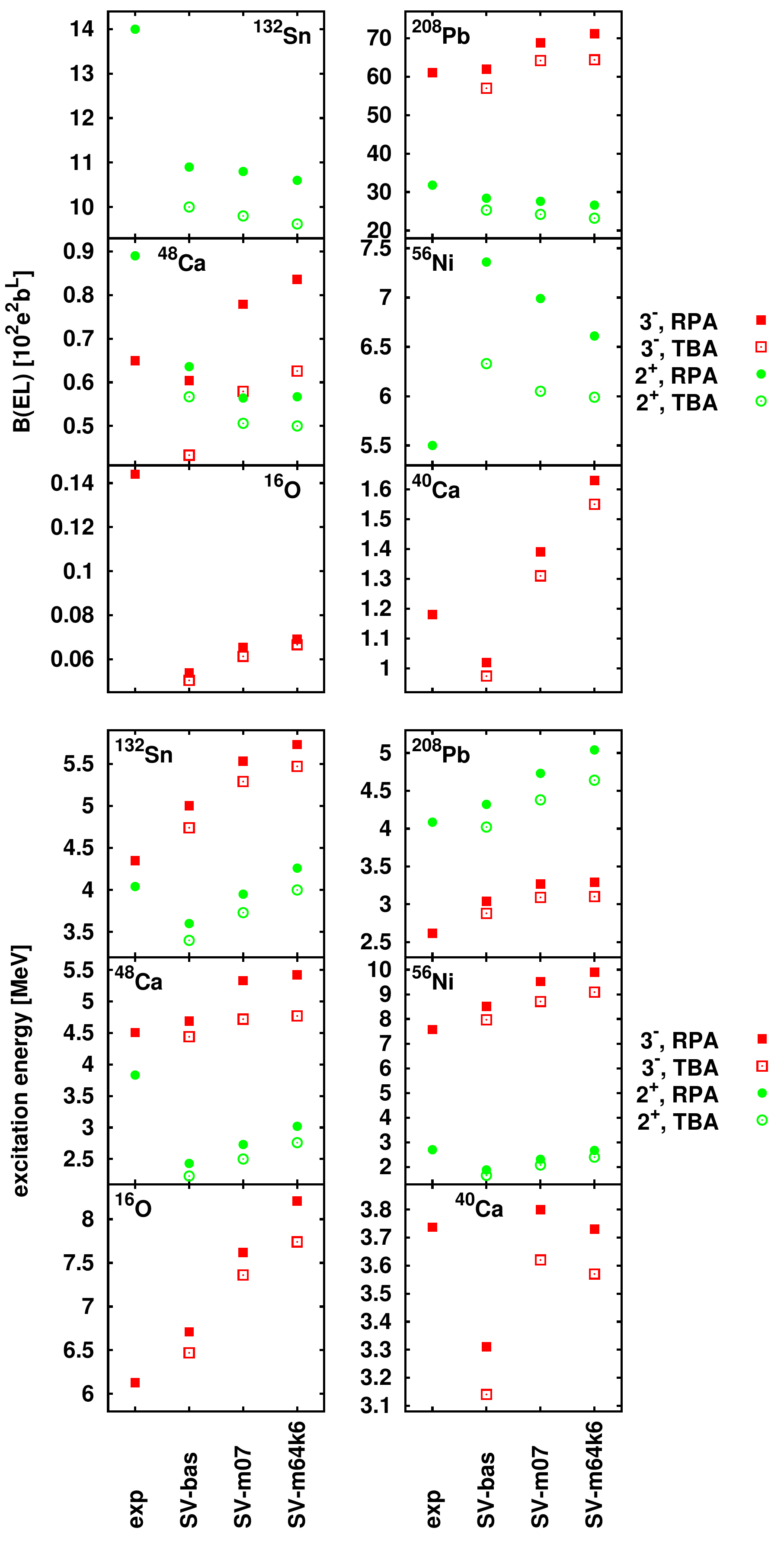}}
\caption{\label{fig:Low-forces} 
Energies and $B(EL)\!\uparrow$ values of low lying $2^+$ and $3^-$ states 
for the three selected Skyrme parametrizations
and six doubly magic nuclei.The experimental values were taken
from the following Refs.:
  $^{16}$O \cite{Tilley_1993},
  $^{40}$Ca \cite{Chen_2017},
  $^{48}$Ca \cite{Burrows_2006},
  $^{56}$Ni \cite{NDS_2011},
  $^{132}$Sn \cite{Khazov_2005},
  $^{208}$Pb \cite{Martin_2007}.
  }
\end{figure}
Fig.~\ref{fig:Low-forces} shows energies and $B(EL)\!\uparrow$ values of the
low-lying collective 3$^-$ and 2$^+$ states in all {doubly} magic
nuclei. The three Skyrme parametrization have different $m^*/m$:
0.9 (SV-bas), 0.7 (SV-m07) and 0.64 (SV-m64k6). As mentioned above,
the low-lying collective $3^-$ state depends on $m^*/m$ in a sensitive
and systematic way, the lower $m^*/m$ the higher $E(3^-)$ because
these energies are dominated by the energy gap between the shells with
opposite parity which increases with decreasing $m^*/m$.  The
parameter set SV-bas gives for all doubly magic nuclei by far the best
results for the energy of the $3^-$ states.  The $E(3^-)$ are
particularly large for $^{56}$Ni. This happens because this is the
first nucleus in the sample where both shells, protons and neutrons,
include the one $sp$ state from the next higher shell which is driven
down by spin-orbit splitting. This effectively enhances the energy gap
between occupied and unoccupied states of opposite parity.  

The energies of the low-lying $2^+$ states although being enabled by
spin-orbit splitting show the same systematic dependence on $m^*/m$
because of their strong coupling with the GQR which depends on $m^*/m$
as seen above.  The theoretical $E(2^+)$
in the medium mass nuclei are slightly too low although the spin-orbit
orbit splitting is rather appropriate as seen from
Table~\ref{Spin-Orbit(2)}. For heavier nuclei, the theoretical value
increases relative to data. In all cases, the deviations are, in fact,
small. In spite of the differences which we probably pointed out too
much, it is astonishing that the Skyrme functionals place the two
low-lying states so well although their energies are not fitted and
emerge from subtle interplay of several ingredients.

The effect of phonon coupling (difference between open and filled
symbols) is generally small (usually 0.2 MeV, occasionally 1 MeV),
smaller than for GR but relative to the lower excitation energies
comparable. The down-shift is usually beneficial for coming closer to
data in case of $E(3^-)$. The situation is mixed for $E(2^+)$ because
the relation of theory to data varies with parametrization and
nucleus. The general conclusions remains the same as for GR: the
general trends are not changed by phonon coupling, but the small
down-shift should be considered when going for fine tuning a model.

\subsection {Transition probabilities $B(EL)$ of the low-lying
collective states}

The transition probabilities, $B(EL)$ values shown in the upper block
of Fig. \ref{fig:Low-forces}, are an even more sensitive test for the
quality of a theoretical approach compared with the energies as they
test the structure of the wave functions.  Fig. \ref{fig:Low-forces}
shows both, energies and $B(EL)$ values, for the low lying 2$^+$ and
3$^-$ collective states together in comparison with data.  The
agreement is satisfactory. Not only the energies but also {$B(EL)$}
values in most cases stay within 10\% to 20\% from data. Only the
$B(E3)$ values in $^{16}$O deviate by a factor of three. This might be
connected with the single-particle energies. It has been shown in
Ref. \cite{Kamerdzhiev_2001} that if one starts with the experimental
$ph$- energies and includes the excitation energy of the 3$^-$ state
in the fitting procedure of the $ph$-force than the $B(E3)$ value is
in good agreement with the data. In the present approach the energy
gap between the particle and hole states is 4 MeV smaller compared
with the experimental values. As the energy of the 3$^-$ is well
reproduced it is obvious that the (self-consistent) $ph$-force is too
weak to create collectivity. To much lesser extend, this can be said
of the other cases too. With the exception of the $B(E2)$ value of
$^{56}$Ni all $B(E2)$-values are already in RPA too small. As mentioned
before, the fragmentation of the single-particle strength gives an
additional reduction in TBA.

In this connection one has to bear in mind that in the present
approach (like in most other approaches of this kind) correlations
beyond RPA are included only in the excited states but not in the
ground state. One of the few exceptions is
Ref. \cite{Kamerdzhiev_1993a}. Here magnetic dipole states were
investigated and in this case, as expected, ground-state correlations
beyond RPA give rise to a reduction of the strength. In an other
investigation \cite{Drozdz_1990} the authors treated $2p2h$
correlations consistently in the excited states as well as in the
ground state. Here giant resonances were investigated and in the case
of the isoscalar the GQR in $^{40}$Ca the peak shows, as expected, an 
enhancement. Unfortunately no results for low-lying collective states
were here reported.

Comparing TBA with RPA we see that phonon coupling always reduces
the $B(EL)$ strength, not much but very systematically. 
This is explained by the fact that, on the one hand, the TBA
excitation energies are shifted down with respect to the RPA ones.
On the other hand, the RPA-IEWSR (inverse energy-weighted sum rule) is conserved
in the TBA with subtraction (see \cite{Tselyaev_2013}) and therefore the
$B(EL)$ values should change in the same direction.
The same coincidence of lowering energy together with lowering $B(EL)$
could be observed already in Fig.~\ref{fig:dependences_208Pb_SVbas}. 

\section{Summary}
\label{summary}

In this publication we present a new selection criterion for phonons
in phonon coupling models. The criterion relies on the inverse-energy
weighted average interaction strength $v_{min}$ in given RPA modes and
so is independent of additional assumptions about external field
operators. It is found to sort out (and discard) unambiguously all
non-collective states. In this way, it restricts in a natural way the
energies and angular momenta of the phonons included in the model
space. As a formal motivation, we could show that our new criterion
may be compared with the macroscopic method where the low-lying
vibrations with the largest deformation parameter give rise to the
strongest coupling to the single-particle states. We investigated
within the framework of the time-blocking approximation (TBA) the
dependence of the results on the magnitude of the new cutoff
parameter $v_{\mbss{min}}$. It is demonstrated that there exists
a plateau where the numerical results depend only weakly on the
parameter $v_{\mbss{min}}$. From this investigations we extract the
quantity $v_{\mbss{min}}$ = 0.05.

We applied the newly tuned scheme to a systematic survey of giant
resonances as well as of the low-lying collective $2^+$ and $3^-$
resonances in light, medium and heavy double magic nuclei where we
used three different Skyrme parametrizations to explore the variances
of predictions.  Thereby we looked at two aspects, first, we studied
the effect of phonon coupling in a wide range of nuclei and modes, and
second, we looked at the performance in comparison to data.  The
phonon coupling in TBA has three effects: a shift of the resonances
peak energies, an enhancement of their width, and a smoothing of the
spectral distributions. The shift is always downward to lower
resonance energies and remains rather small (0.5--1.5 MeV for giant
resonance, 0.2--1 MeV for low lying states). Most importantly, the
down-shift is for a given mode much the same for all nuclei and forces
such that trends (with mass number, with force) are not changed as
compared to RPA. The effect on the width is hardly recognizable as
this is dominated by RPA's fragmentation width.  The largest effect
appears in the detailed spectral distributions which are efficiently
smoothed such that the TBA profiles comes very much closer to the
experimental strength distributions.
  
What agreement of the results with data is concerned, we have
mentioned already the beneficial effect of TBA for spectral
distributions. The only small shift of the peak energies leaves the
burden of matching resonance positions mainly to RPA while delivering
some fine tuning from phonon coupling. And here we find again as has
been worked out in several earlier RPA studies namely that the Skyrme
energy density functional allows a pertinent description of giant
resonances in heavy nuclei but has still unsolved problems with
covering the full $A$ dependence of the collective resonances.
The present survey gives a direction for further search. The problem
of $A$ dependence has first to be resolved roughly within RPA and then
TBA comes into play when fine tuning.

\bigskip

\noindent
Acknowledgment:
\\
This work has been supported by contract Re322-13/1 from the DFG.
N.L. and V.T. acknowledge financial support from Russian Science
Foundation (project No.  16-12-10155).
This research was supported by the Computer Center of SPbU.
We thank Dave Youngblood for providing us with experimental data.

\bibliographystyle{apsrev4-1}
\bibliography{TTT} 

\end{document}